# Beamforming and other methods for denoising microphone array data

Pieter Sijtsma[1]
*PSA3, 8091 AV Wezep, The Netherlands*

Alice Dinsenmeyer[2], Jérôme Antoni[3], Quentin Leclère[4]
*Univ Lyon, INSA-Lyon, F-69621 Villeurbanne, France*

**Measured acoustic data can be contaminated by noise. This typically happens when microphones are mounted in a wind tunnel wall or on the fuselage of an aircraft, where hydrodynamic pressure fluctuations of the Turbulent Boundary Layer (TBL) can mask the acoustic pressures of interest. For measurements done with an array of microphones, methods exist for denoising the acoustic data. Use is made of the fact that the noise is usually concentrated in the diagonal of the Cross-Spectral Matrix, because of the short spatial coherence of TBL noise. This paper reviews several existing denoising methods and considers the use of Conventional Beamforming, Source Power Integration and CLEAN-SC for this purpose. A comparison between the methods is made using synthesized array data.**

## Nomenclature

| | | | |
|---|---|---|---|
| CB | Conventional Beamforming | $g_n$ | steering vector component |
| CSM | Cross-Spectral Matrix | **h** | source component |
| DS | Diagonal Subtraction | $I$ | number of iterations |
| FFT | Fast Fourier Transform | $i$ | imaginary unit |
| MSB | Multiple Source Beamforming | $(i)$ | iteration index |
| PDF | Probability Density Function | $J$ | number of averages |
| PFA | Probabilistic Factor Analysis | $j$ | snapshot index |
| SLRD | Sparse & Low-Rank Decomposition | $K$ | number of sources |
| SNR | Signal-to-Noise Ratio | $k,l$ | source index |
| SPI | Source Power Integration | **L** | PFA mixing matrix |
| SSI | Signal Subspace Identification | $M$ | Mach number of main flow |
| TBL | Turbulent Boundary Layer | $m,n$ | microphone index |
| $A$ | source power [Pa$^2$] | $N$ | number of microphones |
| $a$ | source amplitude | $NI$ | number of integration areas |
| $B$ | auto-spectrum [Pa$^2$] | **n** | noise vector [Pa] |
| **C** | CSM [Pa$^2$] | **p** | benchmark pressure vector [Pa] |
| $C_{mn}$ | cross-spectrum [Pa$^2$] | $r_0$ | reference distance [m] |
| **c** | PFA vector of latent factors [Pa] | $r_1$ | stretched distance [m], Eq. (4) |
| $c$ | speed of sound [m/s] | **s** | signal vector [Pa] |
| **D** | dirty matrix [Pa$^2$] | $s(t)$ | acoustic signal [Pa] |
| **d** | CSM diagonal [Pa$^2$] | $t$ | time [s] |
| $F$ | cost function [Pa$^2$] | **U** | unitary matrix with eigenvectors of **C** |
| **g** | steering vector | $U$ | main flow speed [m/s] |

---

[1] Director, member AIAA; also at Aircraft Noise & Climate Effects, Delft University of Technology, Faculty of Aerospace Engineering, The Netherlands
[2] PhD Student, Laboratoire Vibrations Acoustique; also at Laboratoire de Mécanique des Fluides et d'Acoustique, Univ Lyon, École Centrale de Lyon, France
[3] Professor, Laboratoire Vibrations Acoustique
[4] Assistant Professor, Laboratoire Vibrations Acoustique

| | | | |
|---|---|---|---|
| $U_c$ | convection speed in TBL [m/s] | $\lambda$ | SLRD relaxation parameter |
| **w** | beamforming weight vector | $\mu$ | integration area index |
| $w_n$ | weight vector component | $\vartheta$ | noise multiplier |
| $\vec{x}$ | microphone location [m] | $\sigma(t)$ | noise signal [Pa] |
| $x, y, z$ | cartesian coordinates [m] | $\sigma_0^2$ | accuracy constant in Eq. (29) |
| **α** | PFA vector defining diagonal matrix | $\Upsilon$ | scaling factor |
| $\beta^2$ | $1-M^2$ | $\omega$ | angular frequency [rad/s] |
| $\varphi$ | loop gain | $\vec{\xi}$ | source location [m] |
| $\kappa$ | number of PFA latent factors | $\xi, \eta, \zeta$ | source coordinates [m] |
| $\Lambda$ | diagonal matrix with eigenvalues of **C** [Pa$^2$] | | |

## I. Introduction

THE EU-CleanSky2 project ADAPT is devoted to extracting the acoustic signal from measured data that are contaminated by noise. This typically happens when microphones are mounted in a wind tunnel wall or on the fuselage of an aircraft, and acoustic measurements are severely hindered by hydrodynamic pressure fluctuations in the Turbulent Boundary Layer (TBL).

Denoising the measured data can be done with microphone arrays. Use can be made of the fact that the noise is usually concentrated in the diagonal of the Cross-Spectral Matrix (CSM). This is because, in general, the TBL noise is incoherent between pairs of microphones, implying that the TBL noise cross-spectra tend to vanish in the averaging process. Nevertheless, a residue of TBL noise will remain in the cross-spectra, as the averaging time is always finite.

Several denoising methods have been proposed recently. Some methods[1-3] make use of the fact that the signal part of the CSM must be positive-definite. Other methods[4] exploit the fact that the rank of the signal CSM is usually low. A recent comparison of denoising methods was made by Dinsenmeyer et al[5].

The next step is to utilise the acoustic nature of the signal. That is: only a confined range of wave numbers can be attributed to acoustics. If the locations of the acoustic sources are known, and if the flow between the sources and the microphones is well-described, then the acoustic part of the signal can be extracted by straightforward beamforming. If the source locations are less well-known, and possibly not restricted to isolated locations, then more advanced methods like Source Power Integration[6,7] (SPI), DAMAS[8] or CLEAN-SC[9] may be useful.

Recently, a number of advanced beamforming methods were applied to a benchmark test case of extracting the acoustic signal of a line source from microphone array measurements that were heavily contaminated by incoherent noise[10]. The best results were found with SPI methods, using a narrow integration area enclosing the line.

In this paper, a comparison is made between the following methods:
- Diagonal Subtraction[1-3] (DS),
- Sparse & Low-Rank Decomposition[4] (SLRD),
- Probabilistic Factor Analysis[5] (PFA),
- Conventional Beamforming[9] (CB),
- Source Power Integration[6,7] (SPI),
- CLEAN-SC[9].

The methods are applied to two synthetized measurements with an array of 93 microphones. The beamforming methods (CB, SPI, CLEAN-SC) are applied with and without knowledge of the source positions.

In Section II of this paper the synthetic benchmark data are described. Section III gives a review of the denoising methods. The comparison is described in Section IV and a brief further discussion can be found in Section V. The results are summarized in Section VI.

## II. Array benchmark data

Synthetic benchmark data were generated for an array of 93 microphones, located in the plane $z = 0$. The $(x, y)$ coordinates are plotted in Figure 1. The layout is very similar to the array used by Sarradj et al[10], but without being symmetric about the line $y = 0$.

### A. Noise

For each microphone, 60 s of Gaussian noise $\sigma_n(t)$ was generated, representing TBL noise with an eddy convection speed of $U_c = 60$ m/s in positive x-direction. The auto-spectrum $B(\omega)$, shown in Figure 2, was the same for all microphones. The targeted coherence between pairs of microphones was the Corcos relation[11]. Thus, the cross-spectra between microphones in $(x_m, y_m)$ and $(x_n, y_n)$ should be described by

$$C_{mn}(\omega) = B(\omega) \exp\left\{-\frac{\omega}{U_c}\left(0.116|x_n - x_m| + 0.7|y_n - y_m|\right)\right\}. \tag{1}$$

### B. Signal

In addition, acoustic data were generated for two configurations:
- Case 1: Five omnidirectional sources at 8 m height above the array, at the following positions: $(0,0,8)$, $(-4,-4,8)$, $(-4,4,8)$, $(4,4,8)$, and $(4,-4,8)$.
- Case 2: A line source at 8 m height, represented by 1000 sources between $(0,-1,8)$ and $(0,1,8)$.

Source signals $\sigma_k(t)$ were generated by the same procedure as the TBL noise data, but independently for each source, thus making the sources incoherent. The acoustic transfer from a source in $\vec{\xi}$, with emitted signal $\sigma(t)$, to a microphone in $\vec{x}$ with received signal $s(t)$ is

$$s(t) = a \frac{r_0}{r_1} \sigma\left\{t - \frac{1}{\beta^2 c}\left(-M(x-\xi) + r_1\right)\right\}, \tag{2}$$

with $r_0$ a reference distance and

$$\beta^2 = 1 - M^2, \tag{3}$$

$$r_1 = \sqrt{(x-\xi)^2 + \beta^2\left[(y-\eta)^2 + (z-\zeta)^2\right]}. \tag{4}$$

Herein, $M = U/c$ is the Mach number of the main axial flow. The flow speed $U$ is 85 m/s. The sound speed $c$ is 340.3 m/s.

The source amplitude $a$ in Eq. (2) was used to scale the source levels, as follows:
- For Case 1 the level of each source is 1 dB lower than the level of the previous source, in the above-mentioned order. For Case 2, all sources have equal level.
- The average microphone level is equal to the average TBL noise level. In other words, if $\tilde{a}_k$, $k = 1,..,K$ are the relative source amplitudes, then the absolute amplitudes are $a_k = \Upsilon \tilde{a}_k$, with

$$\Upsilon = \left( N \Big/ \sum_{n=1}^{N}\sum_{k=1}^{K} \tilde{a}_k^2 \frac{r_0^2}{r_1(\vec{x}_n, \vec{\xi}_k)^2} \right)^{1/2}, \tag{5}$$

where $N$ is the number of microphones. The microphone signals are then given by

$$s_n(t) = \sum_{k=1}^{K} a_k \frac{r_0}{r_1(\vec{x}_n, \vec{\xi}_k)} \sigma\left\{t - \frac{1}{\beta^2 c}\left(-M(x_n - \xi_k) + r_1(\vec{x}_n, \vec{\xi}_k)\right)\right\}. \tag{6}$$

### C. Mixed data

The previous sections gave a description of how noise data $\mathbf{n}(t) = (\sigma_1(t),\ldots,\sigma_N(t))^T$ and signal data $\mathbf{s}(t) = (s_1(t),\ldots,s_N(t))^T$ are generated. Benchmark time data $\mathbf{p}(t)$ were generated by combining signal and noise at several Signal-to-Noise Ratios (SNR):

$$\mathbf{p}(t) = \mathbf{s}(t) + \vartheta \mathbf{n}(t). \tag{7}$$

Data at the following SNR-values were generated: −24 dB, −21 dB, −18 dB, −15 dB, −12 dB, −9 dB, −6 dB, −3 dB, 0 dB, +250 dB. The latter effectively contains no noise. The relation between $\vartheta$ and SNR is

$$\text{SNR} = -20\log_{10}(\vartheta) \text{ dB}. \tag{8}$$

The time signals were generated at 50 kHz sampling rate. The time data of the source signals were 4 times over-sampled, because interpolation was required to obtain sampled microphone data.

## III. Denoising methods

This chapter briefly reviews the denoising methods considered in this paper.

## A. Signal Subspace Identification (SSI)

Because the CSM (at fixed frequency) is a positive-definite Hermitian matrix, its singular value decomposition can be written like

$$\mathbf{C} = \mathbf{U}\Lambda\mathbf{U}^*, \tag{9}$$

where $\mathbf{U}$ is a unitary matrix containing the eigenvectors, and $\Lambda$ a diagonal matrix containing the eigenvalues, which are all real-valued and positive. The idea is to split $\Lambda$ in a signal and a noise part:

$$\Lambda = \Lambda_s + \Lambda_n. \tag{10}$$

The noise part of the eigenvalue spectrum (when sorted by magnitude) can be recognized as a slowly varying range, as in the example shown in Figure 3 (blue dots). The signal part is defined by the 5 loudest eigenvalues, which, in the example, clearly protrude above the noise spectrum. Thus, by setting $\Lambda_n = 0$, a denoised CSM is constructed with rank 5:

$$\mathbf{C} = \mathbf{U}\Lambda_s\mathbf{U}^*. \tag{11}$$

For beamforming purposes, this idea was exploited by Sarradj[12], leading to processing time and noise reduction with respect to Conventional Beamforming (CB).

If the SNR is lower than in the example of Figure 3, the breakdown in signal and noise is not so obvious. If, for example, the SNR is 3 dB lower than in Figure 3, the lowest signal eigenvalue doesn't protrude above the noise spectrum anymore. The same holds for the other signal eigenvalues when the SNR is further lowered.

For this type of denoising, it is convenient to know the number of sources (i.e., the rank of the signal CSM), which is not always possible.

SSI can be used in combination with the method described in the next section.

## B. Diagonal Subtraction (DS)

The idea of Diagonal Subtraction is to subtract as much as possible from the diagonal of the CSM, while keeping the resulting matrix positive definite. "As much as possible" is quantified by the maximum sum of the subtracted diagonal elements. In other words, the denoised CSM,

$$\mathbf{C}_{DS} = \mathbf{C} - \text{diag}(\mathbf{d}), \tag{12}$$

is obtained by maximizing

$$F(\mathbf{d}) = \text{tr}(\text{diag}(\mathbf{d})), \tag{13}$$

under the constraints:

$$\begin{cases} 0 \leq d_n \leq C_{nn}, \\ \mathbf{C}_{DS} \text{ is positive-definite.} \end{cases} \tag{14}$$

To solve this optimization problem, two solution procedures were proposed recently. Dougherty[1] proposed to use linear programming to maximize $F(\mathbf{d})$, under the constraint $\mathbf{w}_k^*(\mathbf{C} - \text{diag}(\mathbf{d}))\mathbf{w}_k \geq 0$, for an iteratively increasing set of test vectors $\mathbf{w}_k$. Hald[2] proposed to use the public-domain Matlab Software for Disciplined Convex Programming CVX[13] for solving the problem. Another approach of subtracting noise from the CSM diagonal, while keeping it positive definite, is the "Alternate Projections" method as implemented by Leclère et al[3].

When applied to the ADAPT benchmarks described in this paper, the three algorithms are all quite fast, and yield almost identical denoised spectra. However, the method of Dougherty and the Alternate Projections method did not converge to a positive-definite CSM update. After many iterations, with the Alternating Projections method typically one negative eigenvalue remained, and with Dougherty's method the number of remaining negative eigenvalues was even larger. With the CVX method, the updated CSMs were truly positive-definite. Therefore, Alternating Projections and Dougherty's approach will not be considered further in this paper.

DS is a convenient pre-processor for SSI. This is demonstrated in Figure 3 (red circles). After DS, the noise eigenvalues have decreased considerably, leading to a better separation between signal and noise eigenvalues. The signal eigenvalues have also decreased somewhat, because the signal eigenvalues are polluted by noise too. Thus, after DS, SSI is expected to give more accurate signal levels.

## C. Sparse & Low-Rank Decomposition (SLRD)

Another method for removing the noise from the CSM, proposed by Wright et al[14], is to minimize the following expression:

$$F = \|\mathbf{C}_{\text{SLRD}}\|_* + \lambda \|\mathbf{C} - \mathbf{C}_{\text{SLRD}}\|_1, \quad (15)$$

where $\|\cdots\|_*$ denotes the nuclear norm (sum of the absolute eigenvalues) and $\|\cdots\|_1$ the L1-norm (sum of absolute values of matrix elements). The constant $\lambda$ is a regularization parameter. The idea is to obtain a low-rank signal matrix $\mathbf{C}_{\text{SLRD}}$ and a sparse noise matrix $\mathbf{C} - \mathbf{C}_{\text{SLRD}}$. Further details can be found in Finez et al[4].

In contrast with the DS method, the off-diagonal elements of $\mathbf{C}_{\text{SLRD}}$ are affected as well, compared to the original CSM. Moreover, $\mathbf{C}_{\text{SLRD}}$ does not need to be positive-definite.

The minimization problem, Eq. (15), can be solved with a proximal gradient method, for which a publicly available Matlab function exists[15]. For the regularization parameter, Wright et al suggested $\lambda = N^{-0.5}$, where $N$ is the number of microphones. In our case, with $N = 93$, this is $\lambda = 0.104$.

### D. Probabilistic Factor Analysis (PFA)

A method related to SLRD is the Probabilistic Factor Analysis (PFA), as proposed by Dinsenmeyer et al[5], which uses a probabilistic modelling framework. The objective is to fit the averaged CSM to the following model:

$$\mathbf{C} = \frac{1}{J} \sum_{j=1}^{J} \mathbf{p}_j \mathbf{p}_j^*, \quad (16)$$

with

$$\mathbf{p}_j = \mathbf{L}\,\text{diag}(\boldsymbol{\alpha})\mathbf{c}_j + \mathbf{n}_j, \quad j = 1, \ldots, J, \quad (17)$$

where $\mathbf{c}$ is a vector of latent $\kappa < N$ factors, $\mathbf{L} \in \mathbb{C}^{N \times \kappa}$ is a mixing matrix, $\mathbf{n}$ are the residual errors (independent of the factors) and the index $j$ refers to the snapshot number. The denoised CSM is

$$\mathbf{C}_{\text{PFA}} = \frac{1}{J} \sum_{j=1}^{J} \left( \mathbf{L}\,\text{diag}(\boldsymbol{\alpha})\mathbf{c}_j \right)\left( \mathbf{L}\,\text{diag}(\boldsymbol{\alpha})\mathbf{c}_j \right)^* = \mathbf{L}\,\text{diag}(\boldsymbol{\alpha}) \left( \frac{1}{J} \sum_{j=1}^{J} \mathbf{c}_j \mathbf{c}_j^* \right) \text{diag}(\boldsymbol{\alpha})\mathbf{L}^*. \quad (18)$$

This is the classical factor analysis model, in which is added the diagonal matrix $\text{diag}(\boldsymbol{\alpha})$ to select automatically the minimal model order. Various methods exist to infer the model parameters from the data. Here, we use a Gibbs sampler, a Markov Chain Monte Carlo algorithm, to find the maximum a posteriori estimates of the unknowns[16].

The probability density functions (PDF) assigned to the unknown parameters in model (17) are the following:

| Priors | Hyper-priors |
|---|---|
| $\mathbf{c} \sim \mathcal{N}_{\mathbb{C}}\left(0, \gamma^2\right)$ | $\gamma^2 \sim \mathcal{IG}\left(\mathbf{a}_\gamma, \mathbf{b}_\gamma\right)$ |
| $\mathbf{n} \sim \mathcal{N}_{\mathbb{C}}\left(0, \text{diag}\left(\sigma_n^2\right)\right)$ | $\sigma_N^2 \sim \mathcal{IG}\left(\mathbf{a}_\sigma, \mathbf{b}_\sigma\right)$ |
| $\boldsymbol{\alpha} \sim \mathcal{E}(\mathbf{a}_\alpha)$ | |
| $\mathbf{L} \sim \mathcal{N}_{\mathbb{C}}\left(0, \mathbf{I}/\kappa\right)$ | |

where $\mathcal{E}$ is the exponential PDF, $\mathcal{IG}$ the inverse gamma PDF and $\mathcal{N}_{\mathbb{C}}$ the complex Gaussian PDF. With this method, the user does not have to tune a regularization parameter, but has to choose prior parameters according to his/her knowledge about the noise or signal properties. Here, non-informative priors are considered for the source and noise variances $\gamma^2$ and $\sigma^2$ (i.e., mean value given by the noisy auto-spectra, with high variance) and $\mathbf{a}_\alpha$ following an exponential decay.

This method is computationally expensive, but not very sensitive to its input priors. This denoising technique allows for data reduction and provides a sparse basis that can then be used to reduce the dimension of an imaging problem.

### E. Conventional Beamforming (CB)

If the acoustic signal would be due to a single point source in $\vec{\xi}$, then beamforming is the best option to retrieve the signal. In the frequency-domain, let $\mathbf{g}$ be the "steering" vector describing the point source response at the microphones, for example, in uniform flow:

$$g_n = \frac{r_0}{r_1(\vec{x}_n, \vec{\xi})} \exp\left\{ i \frac{\omega}{\beta^2 c} \left( -M(x_n - \xi) + r_1(\vec{x}_n, \vec{\xi}) \right) \right\}. \quad (19)$$

The CB expression for the source power $A$ follows from minimizing

$$F = \|\mathbf{C} - \mathbf{C}_{\text{CB}}\|_2 = \sum_{m,n} |C_{mn} - C_{\text{CB},mn}|^2. \quad (20)$$

with

$$\mathbf{C}_{\text{CB}} = A\mathbf{g}\mathbf{g}^* \Leftrightarrow C_{\text{CB},mn} = A g_m g_n^*. \tag{21}$$

The $(m,n)$ summation in Eq. (20) does not need to include all $(m,n)$ combinations. For example, the CSM diagonal $(m = n)$ can be excluded. The solution of Eq. (20) is

$$A = \sum_{m,n} w_m^* C_{mn} w_n, \tag{22}$$

with

$$w_n = g_n \bigg/ \left( \sum_{m,n} |g_m|^2 |g_n|^2 \right)^{1/2}. \tag{23}$$

Eq. (22) is the acoustic power at $r_0$ m from the source. It can be shown[17] that the weights defined by Eq. (23) provide maximum noise suppression. The denoised CSM is

$$\mathbf{C}_{\text{CB}} = A\mathbf{g}\mathbf{g}^* = \left( \sum_{m,n} w_m^* C_{mn} w_n \right) \mathbf{g}\mathbf{g}^* = \frac{\sum_{m,n} g_m^* C_{mn} g_n}{\sum_{m,n} |g_m|^2 |g_n|^2} \mathbf{g}\mathbf{g}^*. \tag{24}$$

**F. Multiple Source Beamforming (MSB)**

When more than one source exists, at positions $\vec{\xi}_k$, $k = 1, \cdots, K$, then a denoised CSM can be obtained by multiple application of CB:

$$\mathbf{C}_{\text{CB}} = \sum_{k=1}^{K} \left[ \frac{\sum_{m,n} g_{k,m}^* C_{mn} g_{k,n}}{\sum_{m,n} |g_{k,m}|^2 |g_{k,n}|^2} \mathbf{g}_k \mathbf{g}_k^* \right]. \tag{25}$$

Alternatively, a minimization can be done for all sources simultaneously. Then, we have to minimize

$$F = \|\mathbf{C} - \mathbf{C}_{\text{MSB}}\|_2, \tag{26}$$

with

$$\mathbf{C}_{\text{MSB}} = \sum_{k=1}^{K} A_k \mathbf{g}_k^* \mathbf{g}_k. \tag{27}$$

A standard NNLS-solver[18] can be used to obtain non-negative solutions $A_k$. If the sources are well-separated, that is when

$$|\mathbf{g}_j^* \mathbf{g}_k|^2 \ll |\mathbf{g}_j|^2 |\mathbf{g}_k|^2, \text{ for } j \neq k, \tag{28}$$

then there is not much difference between Eqs. (25) and (27).

Under the following conditions:
- Eq. (28) holds, i.e., the sources are well-separated,
- the minimization, Eq. (20) or Eq. (26), is done without the CSM diagonal,
- the sources are sufficiently far away from the array,
- the signal and noise have a Gaussian PDF,

it can be derived that at least 95% of the denoised average auto-spectrum is obtained within an accuracy of 1 dB if

$$\text{SNR} \geq -10 \log_{10} \left( \frac{N-1}{K} \left[ -1 + \sqrt{1 - \frac{NK}{N-1}(1 - \sigma_0^2 J)} \right] \right), \tag{29}$$

where $J$ is the number of snapshots and $\sigma_0^2 = 0.01334$, which is a number related to the 1 dB tolerance and the 95% confidence. For other values $\sigma_0^2$ will be different. A derivation of Eq. (29) can be found in the appendix.

MSB is actually a special case of SPI, discussed in the section below.

**G. Source Power Integration (SPI)**

The ideal situation of having a few well-separated point sources at known locations does not occur often. Several sources may exist having unknown locations and/or being distributed over some area. In those cases, SPI may be useful. SPI expressions can be obtained by defining groups of scan grid points in the regions where sources are expected. At each point of such an "integration grid" a point source is assumed. These grid sources are incoherent and have equal strength $A$, which is found by minimizing

$$F = \|\mathbf{C} - \mathbf{C}_{\text{SPI}}\|_2, \quad (30)$$

with

$$\mathbf{C}_{\text{SPI}} = A \sum_{k=1}^{K} \mathbf{g}_k \mathbf{g}_k^* . \quad (31)$$

As in the previous section, the L2-norm may be exclusive of diagonal terms. The solution for $A$ is

$$A = \frac{\sum_{m,n} \sum_{k=1}^{K} g_{k,m}^* C_{mn} g_{k,n}}{\sum_{m,n} \sum_{k=1}^{K} g_{k,m}^* \left( \sum_{l=1}^{K} g_{l,m} g_{l,n}^* \right) g_{k,n}} . \quad (32)$$

Analogously to the previous section, the cost function Eq. (30) can be extended to multiple integration areas:

$$\mathbf{C}_{\text{SPI}} = \sum_{\mu=1}^{NI} A_\mu \sum_{k=1}^{K_\mu} \mathbf{g}_{k,\mu}^* \mathbf{g}_{k,\mu} , \quad (33)$$

in which $K_\mu$ is the number of points per integration area and $NI$ the number of areas. The source powers $A_\mu$ can be calculated directly by Eq. (32) or by solving the minimization problem, Eq. (30), by an NNLS-solver[18].

### H. CLEAN-SC

The deconvolution method CLEAN-SC[9] basically decomposes the CSM into a "clean" and a "dirty" part. The "clean" part is built up by "source components" $\mathbf{h}$:

$$\mathbf{C} = \sum_{i=1}^{I} \varphi A_{\max}^{(i)} \mathbf{h}^{(i)} \mathbf{h}^{*(i)} + \mathbf{D}^{(I)} , \quad (34)$$

where $A_{\max}^{(i)}$ is the "dirty map" CB peak level at the $i^{\text{th}}$ iteration step, and $\varphi$ is the "loop gain", $0 < \varphi \leq 1$. The denoised CSM:

$$\mathbf{C}_{\text{CLEAN-SC}}^{(I)} = \sum_{i=1}^{I} \varphi A_{\max}^{(i)} \mathbf{h}^{(i)} \mathbf{h}^{*(i)} \quad (35)$$

is positive-definite.

Just like CB and SPI, CLEAN-SC allows for removing the CSM diagonal. Then, at the first iteration step the diagonal is removed from $\mathbf{D}^{(0)} = \mathbf{C}$. At the following iterations, only the off-diagonal elements are updated, so the diagonal remains zero. The iteration stops when

$$\|\mathbf{D}^{(I+1)}\|_1 > \|\mathbf{D}^{(I)}\|_1 . \quad (36)$$

## IV. Comparison

For each dataset, the CSM was calculated with Welch's method, featuring FFTs on blocks of 500 samples, with energy-preserving Hanning window and 50% overlap, yielding 100 Hz frequency resolution. Hence, the number of averages is 11999. The effective number of averages ("bandwidth-time product") is 6000. Denoised CSMs were calculated with various methods. The denoised auto-spectra, averaged over all microphones were compared to the average signal auto-spectra.

### A. Case 1

For the moderate SNR-value of −6 dB, typical results are shown in Figure 4. The following lines are plotted:
- Signal: The true average auto-spectrum of the signal, in other words, the signal that needs to be recovered. In Figure 4, this line is hidden behind other lines.
- Total: Signal + noise.
- DS: Diagonal Subtraction following the method proposed by Hald[2].
- DS+SSI: Diagonal Subtraction combined with Signal Subspace Identification with 5 sources.
- SLRD: Sparse & Low-Rank Decomposition with relaxation parameter $\lambda = 0.104$.
- PFA: Probabilistic Factor Analysis.

- MSB: Multiple Source Beamforming as defined by Eqs. (26) and (27), with $K = 5$.
- SPI: Source Power Integration obtained with a scan plane of 16×16 m$^2$ with 8 cm mesh size at 8 m height. A division was made into 25 sub-areas ($NI = 25$ in Eq. (33)). The SPI result was calculated separately for each area and summed afterwards.
- CLEAN-SC: obtained with the same scan plane as SPI and loop gain $\varphi = 0.2$.

The steering vectors in the beamforming methods (MSB, SPI, CLEAN-SC) did not include spherical spreading factors. In other words, the results of beamforming are "as measured by the array". The errors, i.e., the differences with the "Signal" line, are shown in Figure 5.

It is observed that all methods perform well or fairly well at frequencies up to 5 kHz. The DS results show a small overprediction of 2 dB. Above 5 kHz, SLRD quickly loses performance. Also SPI loses performance, probably because the main lobes of the sources have become too small compared to the integration areas.

Results with lower SNR-values, ranging from −9 dB down to −24 dB, are shown in Figure 6 to Figure 11. At SNR = −24 dB the DR results are lacking, because of a failure of the cvx-software[13]. All methods show reduction of performance with SNR, but MSB is by far the best. According to the theory presented in the appendix, at least 95% of the MSB errors should remain within 1 dB if SNR ≥ −25.44 dB, in other words, for all benchmark data. This is demonstrated in Figure 12, which is a zoomed version of Figure 11[*]. The small bias observed in Figure 12 is due to the fact that beamforming was done without correction for spherical spreading, in contrast to the generation of the source signals.

The second-best method is PFA, which is remarkable since this method doesn't use knowledge about the acoustic source positions. Up to SNR = −15 dB, the CLEAN-SC and DS+SSI results look still acceptable, within 1 dB error.

The beamforming results (MSB, SPI, CLEAN-SC) in Figure 5 to Figure 11 were obtained using knowledge about the source positions. Either the actual source positions were used (MSB) or a scan plan containing the sources (for SPI and CLEAN-SC). If there is no knowledge about the source positions, far-field beamforming may be attempted. In that case, the steering vectors are based on plane waves. For that purpose, a scan grid was used consisting of 25666 view directions, with approximately 0.9° spacing in two directions (elevation and azimuthal angle). For SPI the number of integration areas is $NI = 111$.

Results for SNR = −15 dB are shown in Figure 13. Obviously, the results with DR, DR+SSI, SLRD and PFA are the same as in Figure 8. As expected, SPI performs worse, as the scan plane does not contain the sources. But, remarkably, CLEAN-SC does not perform worse, but actually a little better. Apparently, far-field beamforming with CLEAN-SC is a good alternative when there is no knowledge about the source positions. Note that the sources are already relatively far (8 m) from the array. Far-field beamforming may not be a good option when the sources are close to the array.

**B. Case 2**

Since Case 2 considers an incoherent line source instead of a number of isolated sources, the rank of the CSM is expected to be full. Therefore, the use of SSI is not possible. Results for SNR = −15 dB obtained with other methods, with the same SPI scan grid at 8 m height, are shown in Figure 14. A special "SPI-line" spectrum is obtained with one group of sources, coinciding with the line ($NI = 1$ in Eq. (33)). The results are comparable to the Case 1 counterpart shown in Figure 8. Only the PFA results have become worse, probably because the signal CSM is now a full-rank matrix.

Figure 15 shows the results obtained with far-field beamforming. As in Figure 13, CLEAN-SC yields the best approximation, but the accuracy is significantly lower now. By looking at other SNR-values it was found that the reduction in accuracy has no relation with the SNR. In other words, at higher SNR the accuracy was comparable to SNR = −15 dB.

## V. Further analysis

Further analysis, not discussed here, showed the following:
- When the grid resolution decreases, the performance of CLEAN-SC decreases too, which is as expected. However, against the expectation, the SPI results get better.
- When the acquisition time is decreased, all methods show a decrease in performance.

---

[*] For SNR = −24 dB it can be shown that there is 95% confidence at 0.73 dB tolerance.

- When the number of microphones is decreased, all results become worse. Only for DS the trend is in the opposite direction.
- When the CLEAN-SC loop gain $\varphi$ is increased, the results become less accurate, except for the case with the line source and far-field beamforming (Figure 15).
- Except for DS and SPI, all methods give good approximations of the cross-spectra.

## VI. Conclusions

The investigations reported in this paper lead to the following conclusions:
- MSB gives the best results, but knowledge about the source positions is required.
- Otherwise, if the number of incoherent sources is low, PFA gives the best results.
- If the number of incoherent sources is high, then CLEAN-SC can be used. The scan distance should be as accurate as possible. However, a useful estimate is also obtained when beamforming is done in the far field (using plane waves).

## Appendix: Broadband noise beamforming error analysis

*Single source*

Consider a microphone array with $N$ microphones. Suppose the measured signal is given by

$$\mathbf{p} = \mathbf{s} + \mathbf{n}, \quad (37)$$

where $\mathbf{s}$ is the $N$-dimensional signal vector and $\mathbf{n}$ represents incoherent noise. If $\mathbf{g}$ is a plane wave steering vector (with $|g_n| = 1$), then the beamforming output is

$$A = \frac{1}{N(N-1)J} \sum_{m,n} \sum_{j=1}^{J} g_m^* \left( s_{m,j} + n_{m,j} \right) \left( s_{n,j}^* + n_{n,j}^* \right) g_n. \quad (38)$$

Herein, the index $j$ refers to FFT time blocks (snapshots) and $J$ is their total number. The $(m,n)$ summation is exclusive of the terms with $m = n$. In other words, beamforming is done without the CSM diagonal.

Now we write for the signal:

$$s_{n,j} = x_j g_n, \quad (39)$$

in which the numbers $x_j$ represent statistical broadband noise variations with expectation value

$$E\left\{ |x_j|^2 \right\} = 1. \quad (40)$$

Then the beamforming output is

$$A = \frac{1}{J} \sum_{j=1}^{J} |x_j|^2 + \frac{1}{N(N-1)J} \left\{ \sum_{m,n} g_n \sum_{j=1}^{J} x_j n_{n,j}^* + \sum_{m,n} g_m^* \sum_{j=1}^{J} x_j^* n_{m,j} + \sum_{m,n} g_m^* g_n \sum_{j=1}^{J} n_{m,j} n_{n,j}^* \right\}. \quad (41)$$

For the expectation value we have

$$E\{A\} = 1. \quad (42)$$

For the variance, we have

$$\sigma^2(A) = E\left\{ |A-1|^2 \right\} = E\left\{ \left( \frac{1}{J} \sum_{j=1}^{J} |x_j|^2 - 1 \right)^2 \right\}$$

$$+ \left( \frac{1}{N(N-1)J} \right)^2 \left[ E\left\{ \left| \sum_{m,n} g_n \sum_{j=1}^{J} x_j n_{n,j}^* \right|^2 \right\} + E\left\{ \left| \sum_{m,n} g_m^* \sum_{j=1}^{J} x_j^* n_{m,j} \right|^2 \right\} + E\left\{ \left| \sum_{m,n} g_m^* g_n \sum_{j=1}^{J} n_{m,j} n_{n,j}^* \right|^2 \right\} \right]. \quad (43)$$

If $x_j$ has a Gaussian probability with unit variance, we can evaluate the first term in the right-hand side of Eq. (43) as

$$E\left\{ \left( \frac{1}{J} \sum_{j=1}^{J} |x_j|^2 - 1 \right)^2 \right\} = \frac{1}{J} E\left\{ \left( |x_j|^2 - 1 \right)^2 \right\} = \frac{1}{\pi J} \int_{-\infty}^{\infty} \int_{-\infty}^{\infty} \left( x^2 + y^2 - 1 \right)^2 \exp\left( -\left( x^2 + y^2 \right) \right) dx dy = \frac{1}{J}. \quad (44)$$

For the expectation values in the second term of Eq. (43) we write

$$\left(\frac{1}{N(N-1)J}\right)^2 \left[ 2(N-1)^2 \sum_{n=1}^{N}\sum_{j=1}^{J} E\left\{x_j^2 |n|_{n,j}^2\right\} + \sum_{m,n}\sum_{j=1}^{J} E\left\{\left|n_{m,j}\right|^2 \left|n_{n,j}\right|^2\right\} \right] = \frac{2(N-1)\Theta^2 + \Theta^4}{N(N-1)J} . \tag{45}$$

Herein, we introduced the RMS-value of the noise:

$$\Theta = E\left\{\left|n_{n,j}\right|^2\right\}^{1/2} . \tag{46}$$

Since the signal was normalized to

$$E\left\{\left|s_{n,j}\right|^2\right\} = 1 , \tag{47}$$

we may consider $\Theta$ as the inverse SNR. We find for Eq. (43):

$$\sigma^2(A) = \frac{1}{J}\left(1 + \frac{2\Theta^2}{N} + \frac{\Theta^4}{N(N-1)}\right) . \tag{48}$$

Assuming $A$ to have a Gaussian probability density, the probability of making an error less than 1 dB is

$$P(\sigma) = \frac{1}{\sigma\sqrt{2\pi}} \int_{10^{-0.1}}^{10^{0.1}} \exp\left(-\frac{(\xi-1)^2}{2\sigma^2}\right) d\xi = \frac{1}{2}\left(\mathrm{Erf}\left(\frac{10^{0.1}-1}{\sigma\sqrt{2}}\right) + \mathrm{Erf}\left(\frac{1-10^{-0.1}}{\sigma\sqrt{2}}\right)\right) . \tag{49}$$

For $\sigma^2 = 0.01334$ we have $P(\sigma) = 0.95$. In other words, if

$$\frac{1}{J}\left[1 + \frac{2\Theta^2}{N} + \frac{\Theta^4}{N(N-1)}\right] \leq \sigma_0^2 = 0.01334 , \tag{50}$$

or, equivalently,

$$\Theta^2 \leq (N-1)\left[-1 + \sqrt{\sigma_0^2 J + \frac{\sigma_0^2 J - 1}{N-1}}\right] , \tag{51}$$

then the probability of making an error less than 1 dB is more than 95%. Suppose, for example, there are $N = 93$ microphones and $J = 6000$ averages. Then Eq. (51) yields $\Theta^2 \leq 735.5$. In other words, SNR $\geq -28.67$ dB.

*Multiple sources*

Now assume there are $K$ incoherent sources, so the following is measured:

$$p_{n,j} = \sum_{k=1}^{K} x_{k,j} g_{k,n} + n_{n,j} . \tag{52}$$

The signal is now scaled through

$$\sum_{k=1}^{K} E\left\{\left|x_{k,j}\right|^2\right\} = 1 , \tag{53}$$

which means that $\Theta$ of Eq. (46) still represents the inverse SNR. Beamforming is done on each source separately. If we assume that

$$\left|\mathbf{g}_k^* \mathbf{g}_l\right| \ll 1 , \tag{54}$$

which means that the sources are not too close to each other and that side lobe levels are low, then the beamforming output can be approximated by

$$A_k = \frac{1}{N(N-1)J} \sum_{m,n}\sum_{j=1}^{J} g_{k,m}^* \left(x_{k,j} g_{k,m} + n_{m,j}\right)\left(x_{k,j}^* g_{k,n}^* + n_{n,j}^*\right) g_{k,n} . \tag{55}$$

The total signal is then estimated by

$$A = \sum_{k=1}^{K} A_k . \tag{56}$$

If the condition of Eq. (55) is fulfilled, then the direct method Eq. (56) gives almost the same results as the inverse method. Again, we have $E\{A\} = 1$. Analogously to Eq. (43) we have

$$\sigma^2(A) = E\left\{\left(\frac{1}{J}\sum_{k=1}^{K}\sum_{j=1}^{J}|x_{k,j}|^2 - 1\right)^2\right\} + \left(\frac{1}{N(N-1)J}\right)^2 \left[E\left\{\left|\sum_{k=1}^{K}\sum_{m,n}g_{k,n}\sum_{j=1}^{J}x_{k,j}n_{n,j}^*\right|^2\right\}\right.$$
$$\left. + E\left\{\left|\sum_{k=1}^{K}\sum_{m,n}g_{k,m}^*\sum_{j=1}^{J}x_{k,j}^*n_{m,j}\right|^2\right\} + E\left\{\left|\sum_{k=1}^{K}\sum_{m,n}g_{k,m}^*g_{k,n}\sum_{j=1}^{J}n_{m,j}n_{n,j}^*\right|^2\right\}\right],$$
(57)

further evaluated to (under the assumption of Eq. (54)):

$$\sigma^2(A) = \frac{1}{J}\left(1 + \frac{2\Theta^2}{N} + \frac{K\Theta^4}{N(N-1)}\right).$$
(58)

The difference with Eq. (48) is that the last term is multiplied with the number of sources, *K*. This term originates from beamforming with noise data. With multiple sources, this needs to be done more often. Analogously to Eq. (51), the 1 dB criterion is now

$$\Theta^2 \leq \frac{N-1}{K}\left[-1 + \sqrt{1 - \frac{NK}{N-1}(1-\sigma_0^2 J)}\right].$$
(59)

With $K = 5$ sources, and everything else the same as above, we obtain $\Theta^2 \leq 349.83$ and $\text{SNR} \geq -25.44$ dB.

## Acknowledgments

This work was performed in the framework of Clean Sky 2 Joint Undertaking, European Union (EU), Horizon 2020, CS2-RIA, ADAPT project, Grant agreement no 754881.

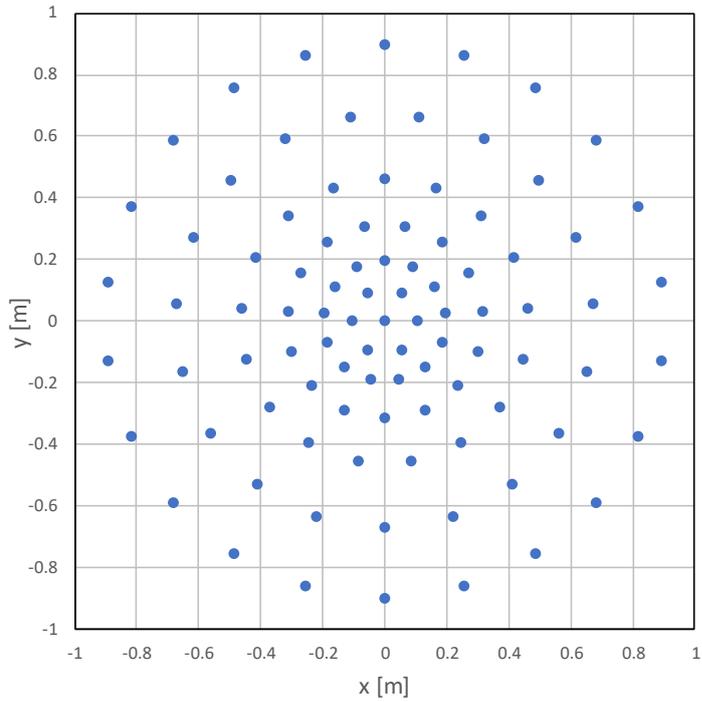

**Figure 1  Array coordinates**

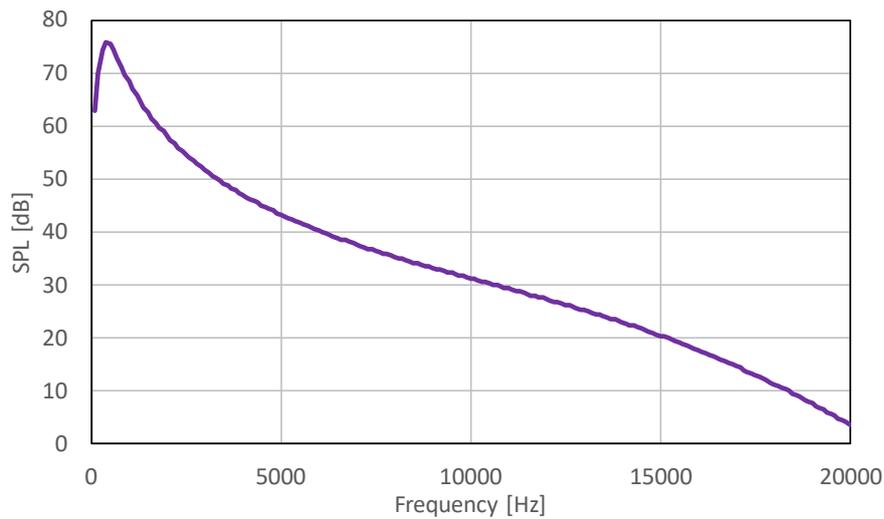

**Figure 2  Microphone auto-spectrum**

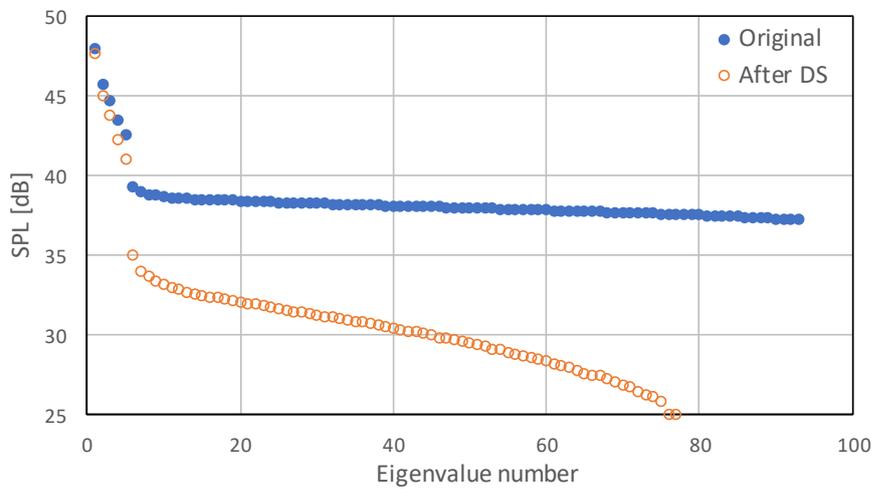

**Figure 3  Eigenvalue spectrum for Case 1, SNR = −6 dB, 3000 Hz; before and after Diagonal Subtraction**

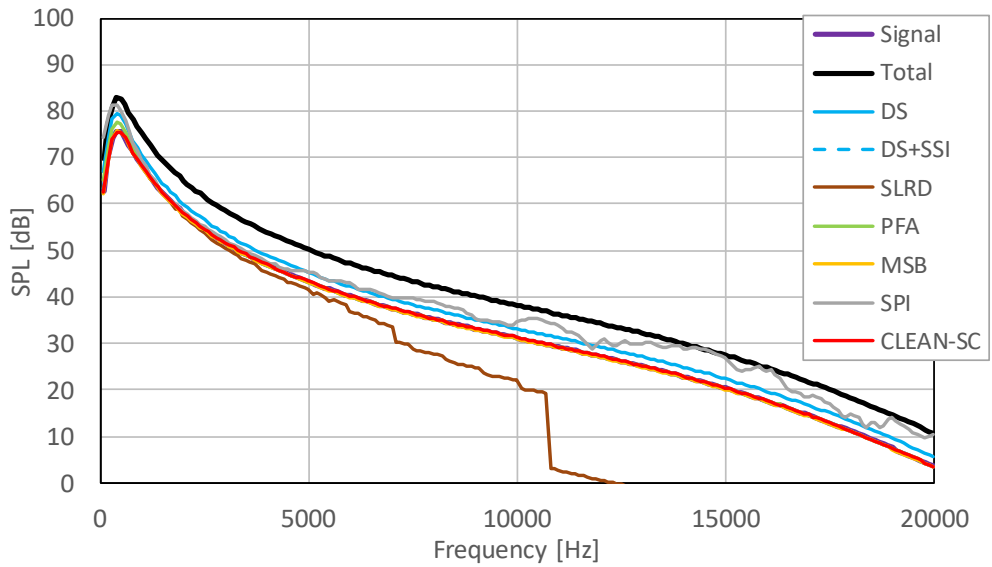

**Figure 4  Auto-spectra obtained with various methods, Case 1, SNR = −6 dB**

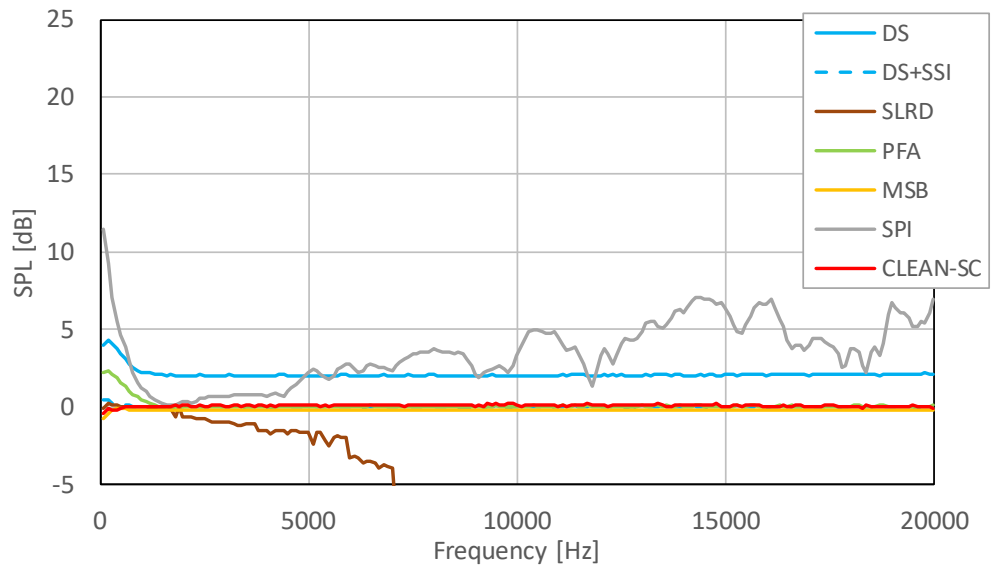

**Figure 5  Errors in auto-spectrum estimations, Case 1, SNR = −6 dB**

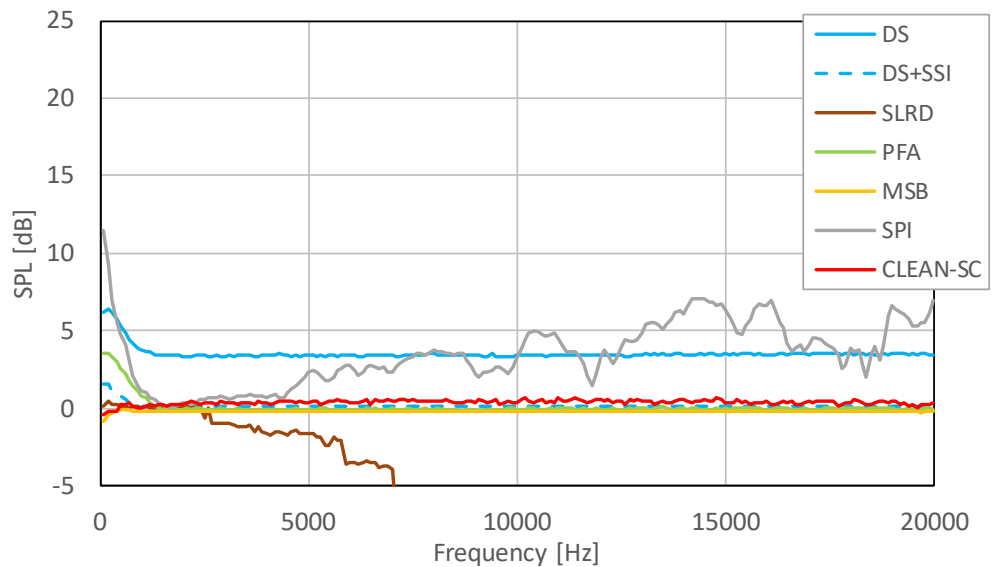

**Figure 6  Errors in auto-spectrum estimations, Case 1, SNR = −9 dB**

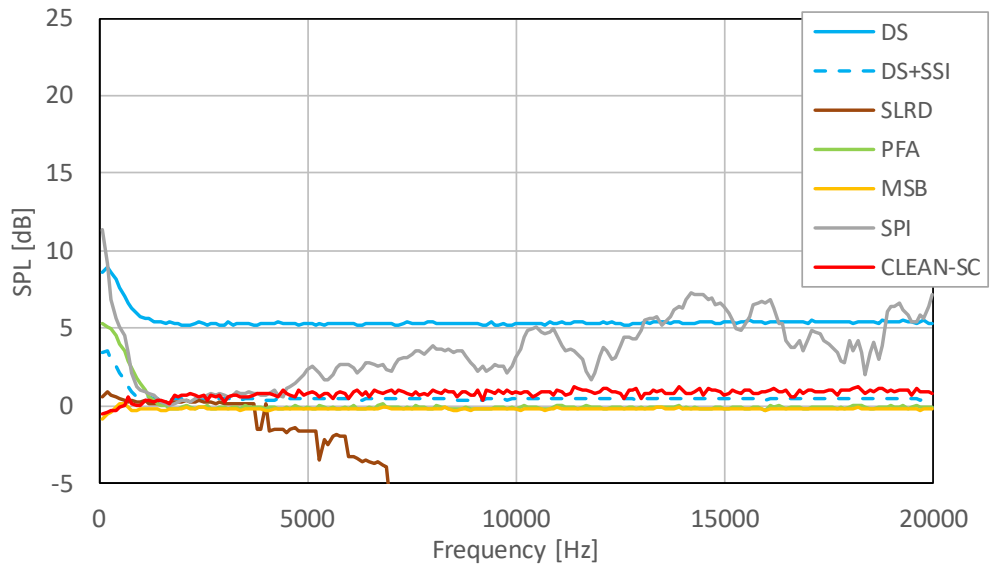

**Figure 7  Errors in auto-spectrum estimations, Case 1, SNR = −12 dB**

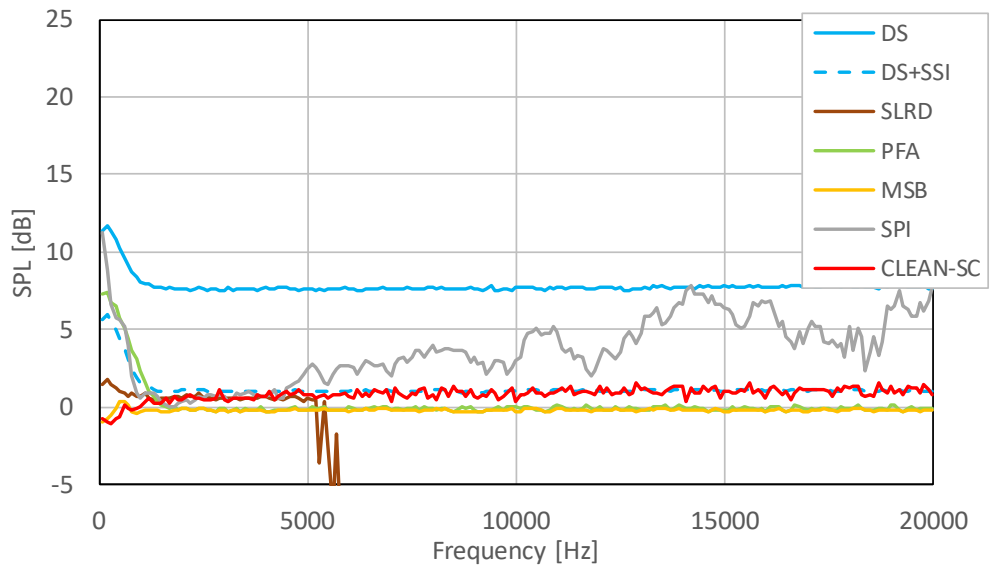

**Figure 8  Errors in auto-spectrum estimations, Case 1, SNR = −15 dB**

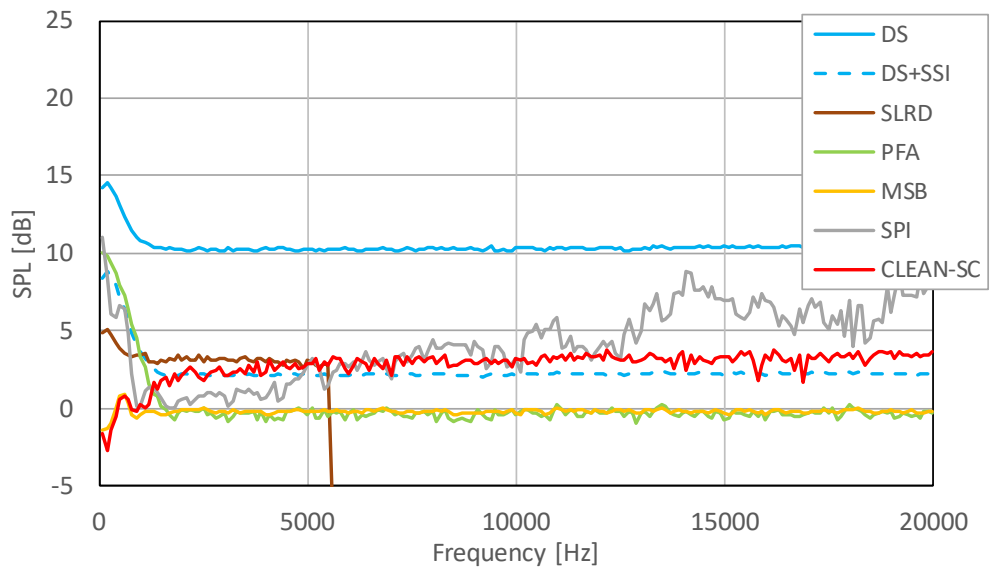

**Figure 9  Errors in auto-spectrum estimations, Case 1, SNR = −18 dB**

14
American Institute of Aeronautics and Astronautics

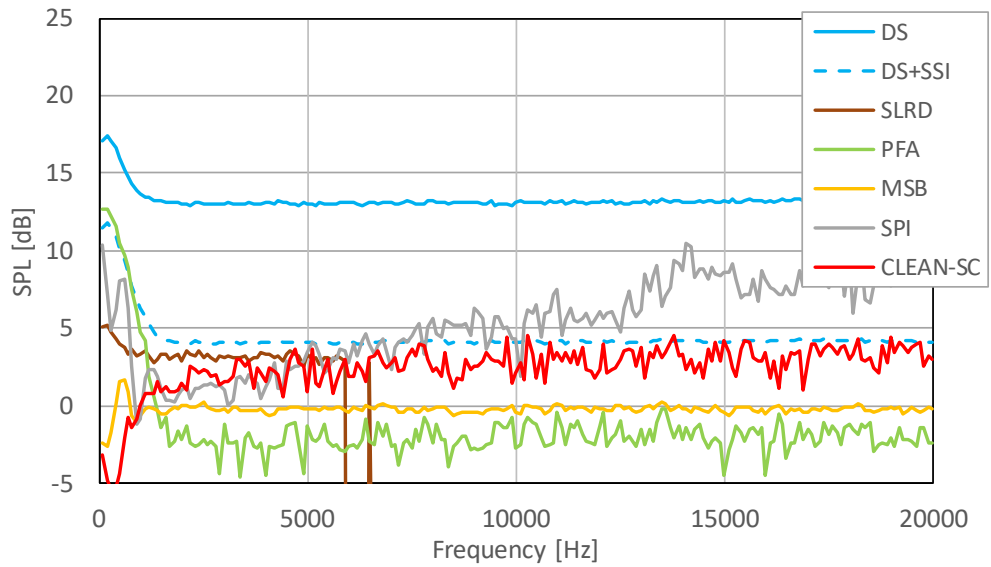

Figure 10  Errors in auto-spectrum estimations, Case 1, SNR = −21 dB

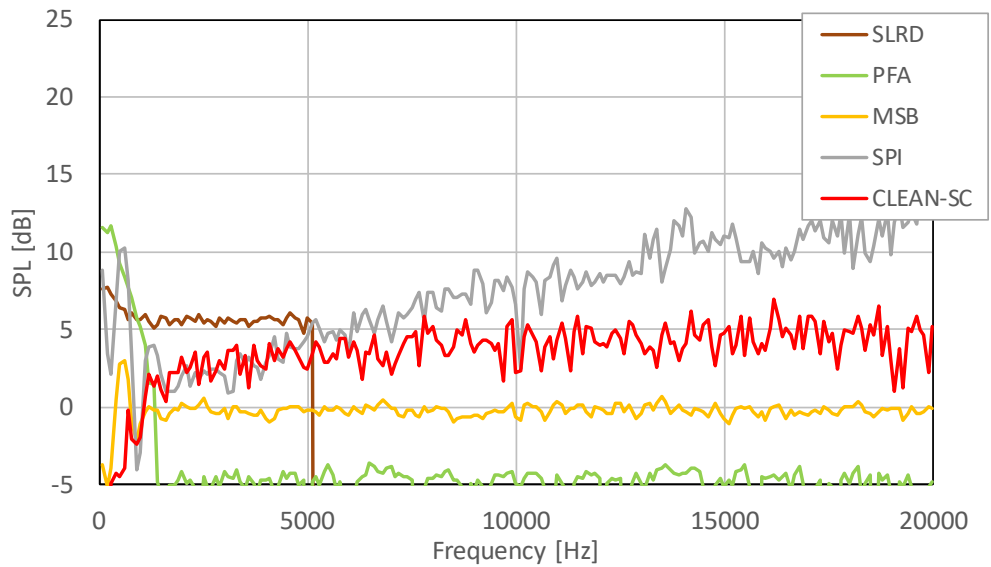

Figure 11  Errors in auto-spectrum estimations, Case 1, SNR = −24 dB

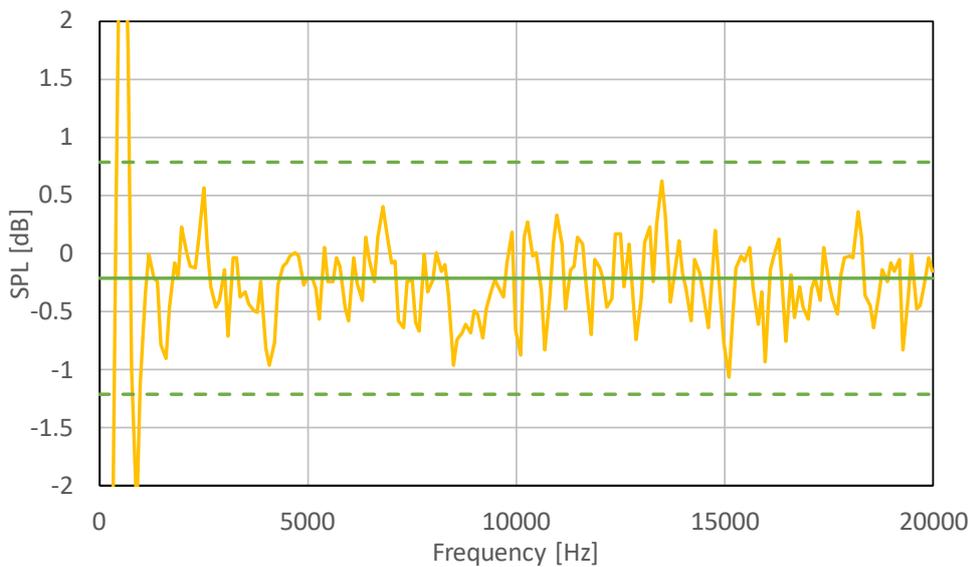

Figure 12  MSB errors in auto-spectrum estimations, Case 1, SNR = −24 dB

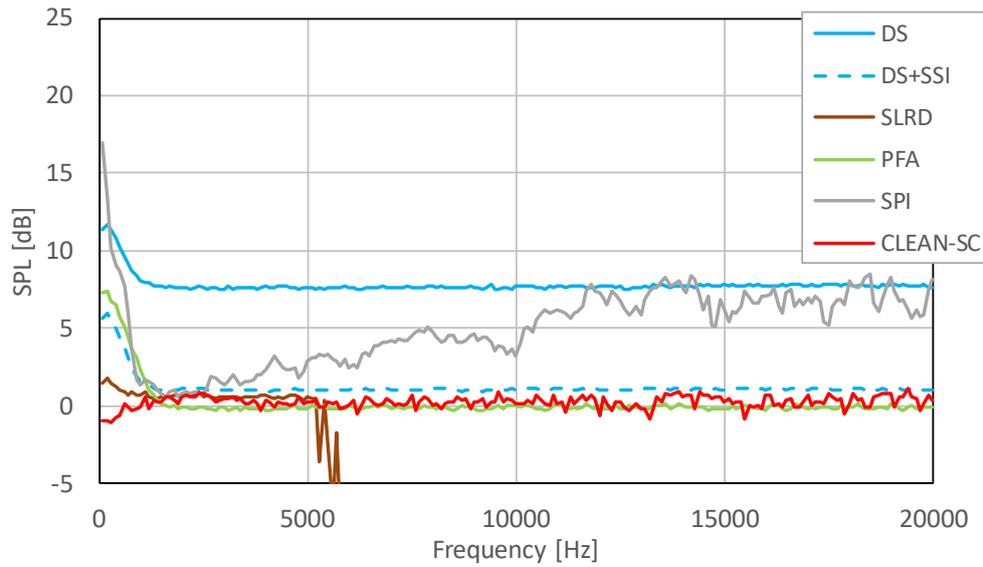
**Figure 13  Errors in auto-spectrum estimations, Case 1, SNR = −15 dB, far-field beamforming**

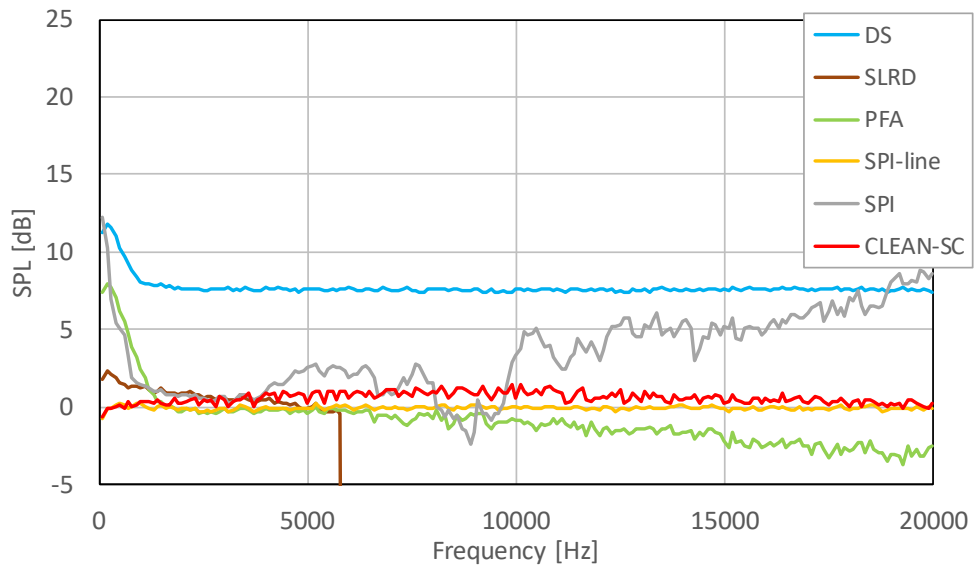
**Figure 14  Errors in auto-spectrum estimations, Case 2, SNR = −15 dB**

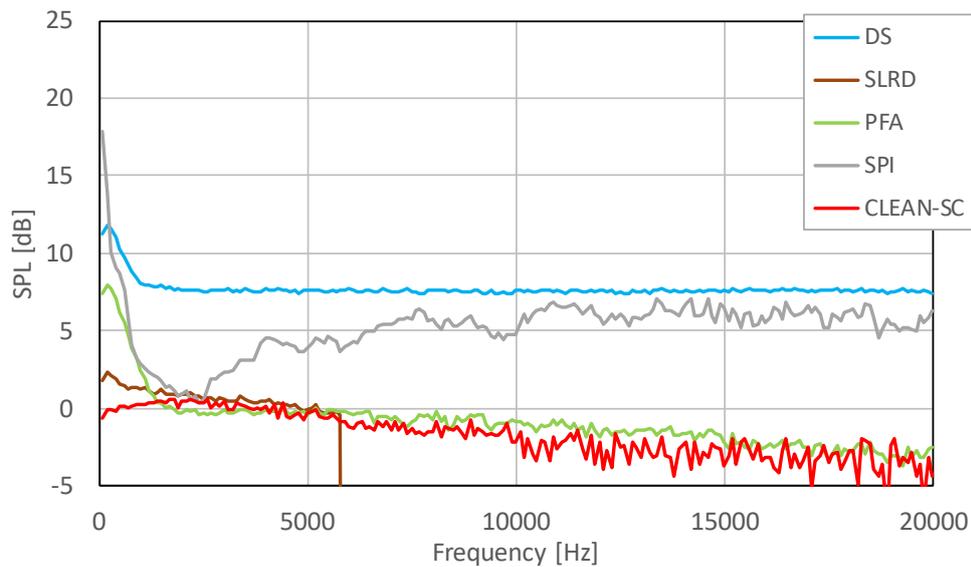
**Figure 15  Errors in auto-spectrum estimations, Case 2, SNR = −15 dB, far-field beamforming**